\documentclass[review,number,sort&compress]{elsarticle}
\usepackage{lineno}

\usepackage{graphicx}
 \usepackage{amsmath}
 \usepackage{multirow}
 \usepackage{url}

\begin{document}

\begin{frontmatter}

\title{Matching problem for primary and secondary signals in dual-phase TPC detectors}
\author[mymainaddress]{B. Radics\corref{mycorrespondingauthor}}
\cortext[mycorrespondingauthor]{Corresponding author}
\ead{bradics@phys.ethz.ch}

\author[mysecondaddress]{E. Burjons}
\author[mymainaddress]{A. Rubbia}
\address[mymainaddress]{ETH Z\"urich, Institute for Particle Physics and Astrophysics, CH-8093 Z\"urich, Switzerland}
\address[mysecondaddress]{ETH Z\"urich, Department of Computer Science, CH-8092 Z\"urich, Switzerland}

\begin{abstract}
The problem of matching primary and secondary light signals, belonging to the same event, is presented in the context of dual-phase time projection chambers. In large scale detectors the secondary light emission could be delayed up to order of milliseconds, which, combined with high signal rates, could make the matching of the signals challenging. A possible approach is offered in the framework of the Stable Marriage and the College Admission problem, for both of which solutions are given by the Gale-Shapley algorithm.
\end{abstract}

\end{frontmatter}


\section{Introduction}
Time projection chambers (TPC) using liquid scintillator are used for three dimensional event reconstruction \cite{CRubbia}. Liquid TPC detectors operate with an electric field across the liquid, and as a result of an ionisation by a primary particle, liberated electrons can be drifted up to the order of meters. A variant of the liquid TPC detectors is the dual-phase (DP) TPC, which allows for additional charge amplification of the drifted electrons by employing a strong electric field across a small gas gap above the liquid phase \cite{ARubbia}. This realisation produces primary and secondary signals, the latter being delayed due to the finite drifting of electrons in the liquid phase.
Measuring the time of delay of the secondary signal encodes information on the drift distance, and therefore provides the position of the event. 

The electron drift velocity, however, may be on the order of $\sim$ mm/$\mu$s, leading to a delay of the secondary light signal up to the order of milliseconds for large scale liquid scintillator experiments, which complicates the reconstruction of the events. Because the sensitivity of the measurements scales with the active target size the upcoming DP TPC detectors will have sizes in the order of several meters, which automatically introduces the problem of how the matching of the primary and secondary light signals can be performed for single or multiple scatter events. Experiments that may face such scenario are typically based on large scale liquid scintillator installation such as the DUNE far detector \cite{DUNE1,DUNE2}, DarkSide-20k \cite{DS20k}, and ArDM \cite{ArDM}.

In this work we provide a possible approach to the DP TPC event matching problem using an  algorithm from Gale and Shapley \cite{GaleShapley}, which gives a solution to the stable marriage problem. Matching, in mathematical sense, is selecting a set of independent edges in a graph without common vertices \cite{GraphTheoryWest}. When applying the concept to DP TPC data, vertices are assumed to be the events and edges are the possible combinations of the events. The problem is to find matching between two classes of events (primary and secondary signal events) in such a way that there are no two events from one class, which could both belong to the same event from the other class. For single scatter events, in which a particle deposited energy only once in the detector, this matching scheme can be applied. For multiple scatter events, other algorithms may also be considered as there could be additional secondary events belonging to the same primary event. Gale and Shapley also provided a solution to a similar problem, the so-called college admission problem \cite{GaleShapley}. In this paper, we study the application of the stable marriage and college admission algorithms both to single and multiple scatter events using data generated by a toy Monte Carlo.

\section{\label{sec:SMP_and_CAP_Problems}The stable marriage and the college admission problems}
\subsection{\label{sec:StableMarriageProblem}The stable marriage problem}
The problem of matching prompt and delayed secondary signals is similar to that of assigning members of a group to another group based on the individual ranking of the group members. The context of the paper of Gale and Shapley is how to reach stable marriages (only in mathematical sense) between a set of men and women, each having their own ranking for the opposite gender candidates. The algorithm consists of an iterative procedure which stops when an actual stable set of marriages is found. The men first propose to the first ranking women on their list, and each woman rejects all but her favourite from those who proposed to her. The women then keep their selected candidate on a string to allow the possibility of a proposal from higher ranking men on their list in the next iteration. The rejected men will propose in the following iteration to the next candidate on their rankings. The same proposal/rejection routine is repeated iteratively until all women have been proposed to. The algorithm is stable, that is by the end there is no such pair of unmatched man and woman who would prefer each other over their previously established candidates. 
This algorithm and its adaptations have been successfully applied in the assignment of medical students to universities, job assignments, roommate
selections among others \cite{GraphTheoryWest}. Variations include: unweighted graphs, where there is 
no preference list for the candidates, assigning multiple pairings to
one vertex (college admission, see below) and matching in non bipartite graphs, where there
is only one type of data (roomate selections)\cite{GI89,Knuth76}.
The criteria for a good solution
might also vary. For example, the Hungarian algorithm \cite{Kuhn55,Mun57} finds the matching with the 
highest likelihood even if it is not stable.
This algorithm and its variations have also been sucessfully implemented and its performance
improved.
All of these together with other approaches appear in the books by Gusfield-Irving \cite{GI89} and Knuth \cite{Knuth76}.

\subsection{\label{sec:CollegeAdmissionProblem}The college admission problem}
The use case of matching multiple members of a group to single members of another group is covered by the solution to the college admission problem. Briefly, students apply to a certain number of colleges, each with certain quotas and rankings for the students, while each student having their own ranking for the colleges. In the first iteration of the solution, students apply to the first colleges on their ranking lists, and the colleges take a number of students according to the available quotas, and put these students on a waiting list, while the rest of the students are rejected. In the next iteration, the previously rejected students apply to the second colleges on their ranking lists. The colleges consider the new applicants and compare them with those on their waiting lists, picking only the top students from the two sets according to their ranking lists, while rejecting the rest. The iteration terminates when every student is on a waiting list or has been rejected from all colleges. At this point all students on the waiting lists are admitted to the colleges.

\section{\label{sec:DelayedSignalProblem}The delayed secondary signal matching problem with single scatter events}
\begin{figure}[htb]
\centering
\includegraphics[height=0.5\textwidth]{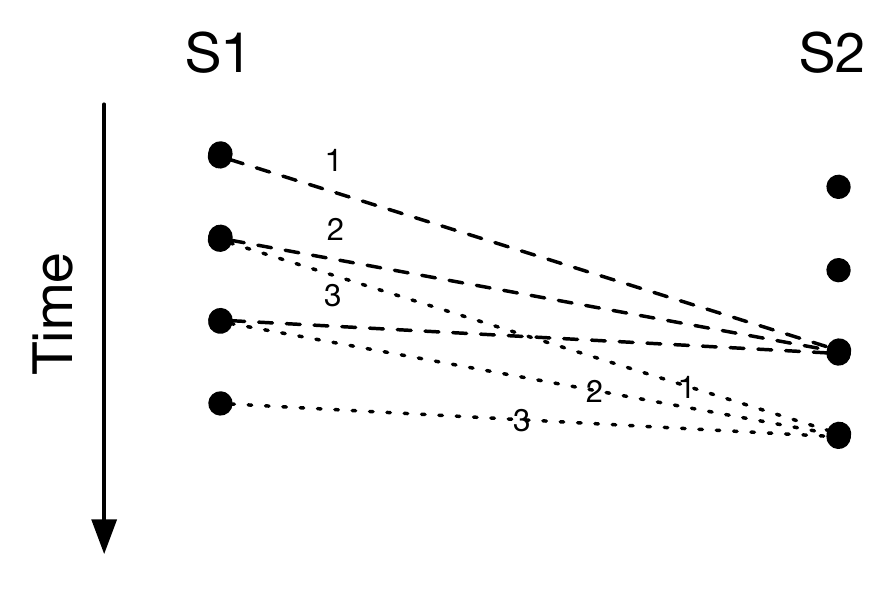}
\caption{Illustration of matching possibilities of primary (S1) and secondary (S2) events to each other. The dashed and dotted lines indicate candidate edges for fixed S2 events to multiple S1 candidate events. The numbers on the edges indicate possible ranking orders which is to be estimated from detection specific information.}
\label{fig:graph}
\end{figure}

The context presented can be translated into the problem of matching primary and delayed secondary signals from the same event. The situation is illustrated in Fig.~\ref{fig:graph}.
Events are recorded sequentially in time order and are assumed to be classified to be primary (S1) or secondary (S2) signal type. Upon detecting a secondary signal event a ranking list can be constructed, containing a value for each of a set of previously detected primary signal events such that the first in the ranking list is the most compatible to match the secondary signal to. The ordering rule in the ranking must be based on some information commonly shared by the event characteristics (such as event topology, etc.). The Gale-Shapley algorithm gives a stable solution given any plausible ranking table, but it is not necessarily the true solution (experimental conditions, such as attenuation of drifting charge or inefficient extraction of electrons from the liquid phase, might be present in the data). The ordering rule plays a key role in finding the best matching. In addition to a experimentally validated ordering rule, the primary and delayed secondary signal candidates must also be present in the data set analysed, which might put some constraints on the size of the dataset used to search for the matching signal.

In any case, assuming the above conditions are met, an ordering rule can always be constructed (e.g. as a Likelihood function), which gives a measure of compatibility between signal events. In this way a ranking list can be made for each primary (secondary) signal events with respect to all later (earlier) opposite type events. In the following we use a Likelihood ordering and illustrate a simplified situation of toy Monte Carlo simulated primary and delayed secondary signals.

\subsection{\label{sec:Problem}Application to single scatter toy Monte Carlo simulation}
Here we assume a simplified scenario of a DP TPC with 1.5 m drift length, electron drift velocity $v_{\mathrm{drift}} = 1.3$ mm/$\mu$s.  The primary signals (S1) are generated randomly and uniformly across the 1.5 m long detector, and for each primary signal, the secondary signal (S2) is generated after a delay time given by the electron drift velocity and the distance of the primary event from the liquid surface. The position of the primary event may be smeared by a Gaussian in order to mimic the detector response. 
In the following, a Likelihood function is constructed in order to perform the ranking of events for matching. Most experiments can recover some rough event topology information from the relative amount of light detected in various subdetectors. We assume this is the case also in this simplified scenario, which means that a rough guess can be made on the position, $x_{\mathrm{s1}}$, of the primary signal event in the liquid. Therefore having a particular secondary signal event, S2, detected for each previously detected primary signal event, S1, the Likelihood of being the correct match can be calculated formally as,
\begin{equation}
L(S1| S2) \propto g(x_{\mathrm{s1}}, t_{\mathrm{s1}}|  x_{\mathrm{s2}}, t_{\mathrm{s2}}, v_{\mathrm{drift}}) , t_{\mathrm{s1}} < t_\mathrm{s2}
\label{eq:Likelihood}
\end{equation}
In the above formulation, $g$ denotes an arbitrary measure of probability, but for a simplified case a Gaussian is assumed, whereby signal pairs with calculated delay time $dt' = (x_{\mathrm{s2}} - x_{\mathrm{s1}})/v_{\mathrm{drift}}$ closer to the observed values of $dt =  t_{\mathrm{s2}} -  t_{\mathrm{s1}}$ get a higher probability. 

As a toy example, 5 events are presented for primary and secondary signals, generated with 1 kHz fictitious rate for the primary signal. Events are shown in order of the time of their detection in Table~\ref{ex_tabEvents}. In the table, $x$ indicates the calculated position for the event from the subdetector information, and for each S2 delayed secondary signal the corresponding true S1 primary event is also indicated. The generated events demonstrate the property that sometimes a primary signal might be detected before the secondary signal arrived for the previous primary event. 
\begin{table}[ht]
	\begin{tabular}{l | l | l | l } \hline \hline
\bf{Event}   &  time [ms] & $x$ [cm] &  Type\\ \hline
1 &1 & 137.13974  & S1 \\
2 &1.10486 & 143.64141  & S2 (1)\\
3 &2 & 27.871302  & S1 \\
4 &3 & 58.513219  & S1 \\    
5 &3.1447 & 139.29184  & S2 (3)\\
6 &3.69512 & 181.66746 &  S2 (4)\\ 
7 &4 & 94.253070  & S1 \\
8 &4.63494 & 150.30378  & S2 (7) \\
9 &5 & 6.1128181  &  S1  \\
10 &   5.87869 & 175.79793  & S2 (9)\\
\hline \hline
	\end{tabular}  
	\caption{Example toy Monte Carlo signal events generated sorted in time order. Primary signal events (Events 1, 3, 4, 7 and 9) have been generated at each millisecond, the delayed secondary signal events (Events 2, 5, 6, 8 and 10) have been detected with the delay time given by the electron drift velocity.  The type of event (S1: primary, S2: secondary) is indicated in the last column, and the true primary event number is indicated for each secondary event.}
	\label{ex_tabEvents}
\end{table}

The ranking order for these events has been calculated using the Likelihood function given in Eq.~\ref{eq:Likelihood}. The corresponding ranking tables are presented in Table~\ref{ex_tabS1} and Table~\ref{ex_tabS2}, respectively. The condition that the delayed secondary signal must happen, by construction, later than the primary signal is explicitly visible in the empty cells in the ranking tables. In the example we gave full ranking tables, however, in practice the knowledge of the maximum delay time may put further constraint on the matching candidates.
\begin{table}[ht]
	\begin{tabular}{l | l | l | l | l | l} \hline \hline
	  \bf{S1 time [ms]} & Rank1 & Rank2 & Rank3 & Rank4 & Rank5 \\ \hline 
	   1  & 2	& 5	&6	&8	&10\\
	   2 & 5&	6&	8&	10 & - \\
	   3 & 6&	5&	8&	10&-\\
	   4 & 8	&10	&-&	-&	-\\
	   5 & 10	&-&	-&	-&	-\\
	   \hline \hline
	\end{tabular}  
	\caption{Ranking table for primary signal candidates, S1. For each row the columns are in order of ranking for the probably matching secondary signal events, S2, indexed by their event number. }
	\label{ex_tabS1}
\end{table}
 
\begin{table}[ht]
	\begin{tabular}{l | l | l | l | l | l} \hline \hline
	  \bf{S2 time [ms]} & Rank1 & Rank2 & Rank3 & Rank4 & Rank5 \\ \hline 
	   1.10486  & 1&	-&	-&	-&	-\\
	   3.1447 & 3&	4&	1&	-&	-\\
	   3.69512 & 4	&3	&1	&-	&-\\
	   4.63494 & 7&	 4&	3&	1&	-\\
	   5.87869 & 9&	7&	4&	3&	1\\
	   \hline \hline
	\end{tabular}  
	\caption{Ranking table for secondary signal candidates, S2. For each row the columns are in order of ranking for the probably matching primary signal events, S1, indexed by event number. }
	\label{ex_tabS2}
\end{table}

The application of the Gale-Shapley iterative matching algorithm thus assigns the events from the two groups to each other, given the information from the ranking tables, and ultimately from the Likelihood function. The performance of the algorithm is studied in the next section. 

It is noted that the size of the ranking table may depend on the particular problem and on the constraints in the data. The example was given only to illustrate the framework of the algorithm.

\subsection{\label{sec:PerformanceFullInfo}Performance of the Stable Marriage algorithm}
We distinguish two cases in order to study the performance of the match making: events with full information and events with various amounts of position smearing. In case the Likelihood function contains full information it should place the correctly matched events first in the ranking table. However, for certain combinations of event rates and delay times it may occur that by chance a pair of uncorrelated primary and secondary events gets a better position in the ranking table than the true pairs. 

In order to search matching candidate for a given event we used a time window whose size is twice the size of the expected maximum delay time for the given detector volume. The reasoning behind is that during constructing the ranking table for a fixed S2 event out of earlier S1 events, the latest S1 event in the search window (although a priori not known) most likely has produced a different S2 event at a later time than the original one, maximum after another full length of drift time.

\begin{figure}[htb]
\centering
\includegraphics[height=0.7\textwidth]{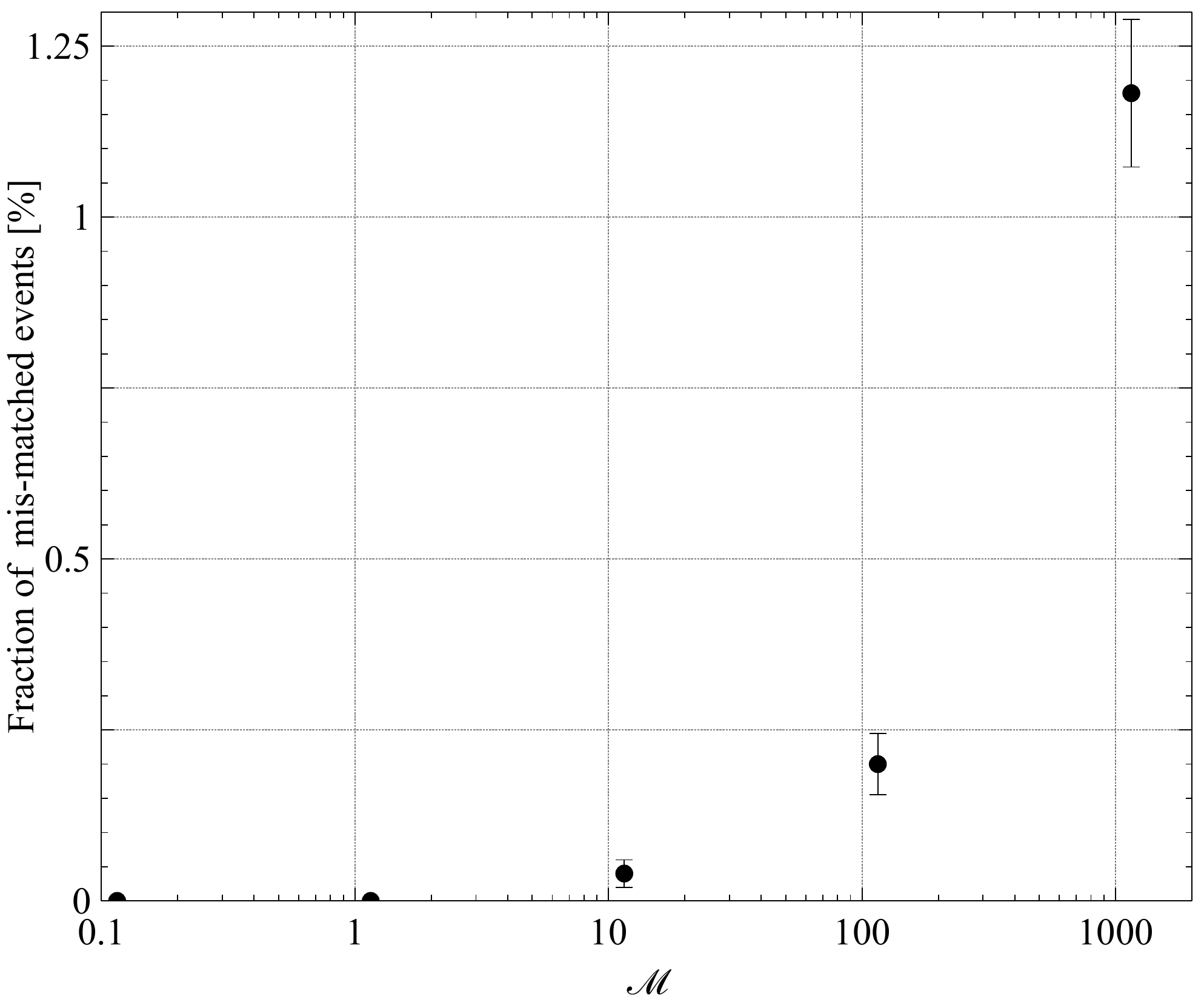}
\caption{Fraction of mis-matched events in the Stable Marriage algorithm as a function of the dimensionless quantity $\cal{M}$. For each point $10k$ events were generated. }
\label{fig:Bad_Full_M}
\end{figure}

We quantify the performance of the algorithm by reporting the fraction of mis-matched events. We define the dimensionless quantity, $\cal{M} = \cal{R} \times \cal{T}$, where $\cal{R}$ is the event rate and $\cal{T}$ is the delay time. Lower values of $\cal{M}$ indicates lower chance of random mismatching. For each value of $\cal{M}$ we generated $10k$ events and have run the reconstruction algorithm with full information in the Likelihood used for ranking. Figure~\ref{fig:Bad_Full_M} shows the fraction of mis-matched events as a function of $\cal{M}$. The results show that even above an extreme value of $\cal{M} \simeq$ $1000$ (equivalent of $\cal{R} \simeq $ $10^{6}$ Hz rate and delay time of $\cal{T} \simeq$ $1$ ms) the fraction of mis-matched events is still at the level of $1 \%$, however, increases exponentially with the dimensionless quantity, $\cal{M}$.  

\begin{figure}[htb]
\centering
\includegraphics[height=0.7\textwidth]{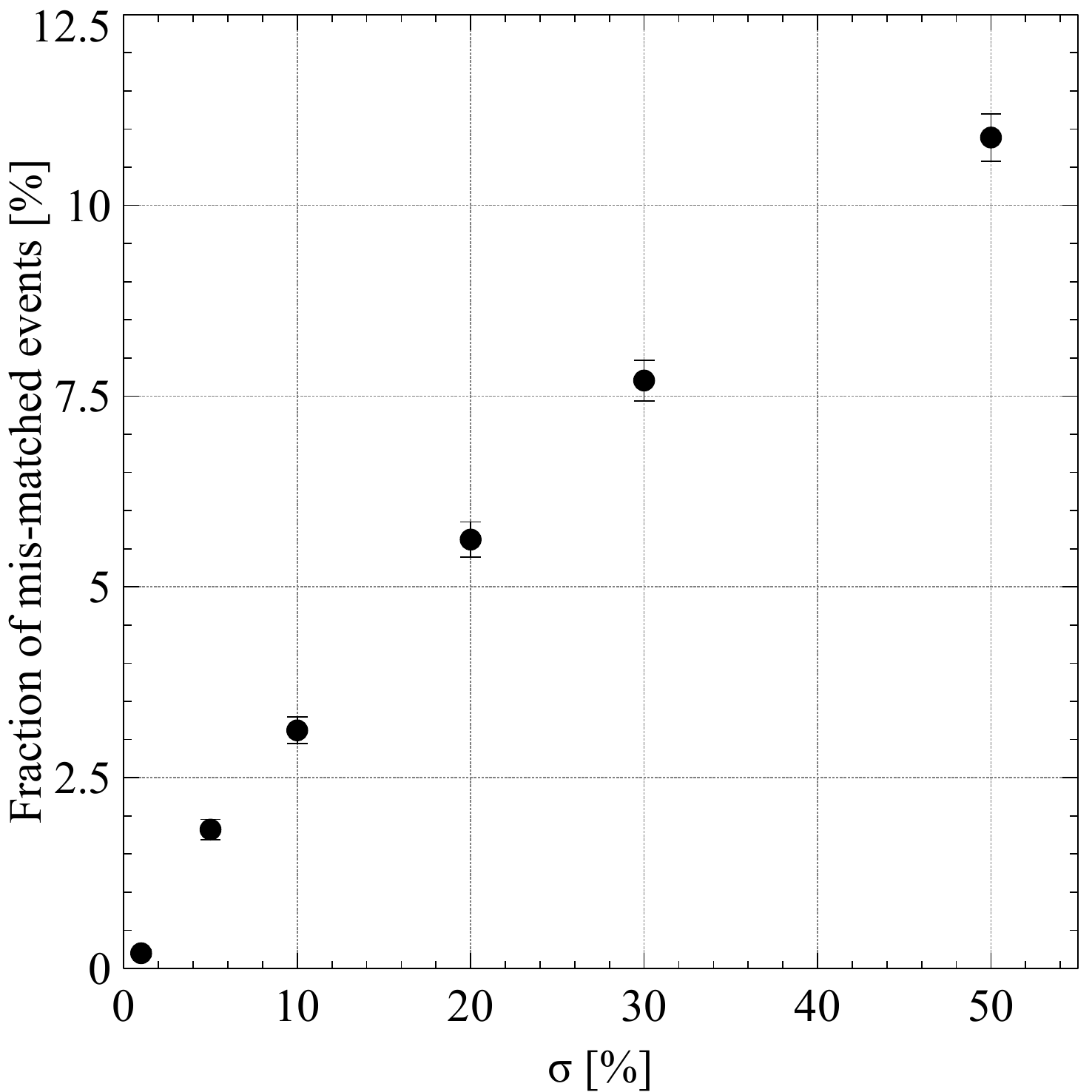}
\caption{Fraction of mis-matched events in the Stable Marriage algorithm as a function of various relative amounts of Gaussian smearing, $\sigma$, on the primary signal emission position, in units of percent, for events with partial information in the Likelihood. For each point $10k$ events were generated.}
\label{fig:Bad_Part_S}
\end{figure}

The performance evaluation is repeated with application of certain amount of position smearing on the primary signal position, but fixing $\cal{M}$ to $\cal{M} \simeq$ $ 1$, in order to minimize the cases of random mismatching. The match making reconstruction algorithm is applied and again the fraction of mis-matched events is reported. Figure~\ref{fig:Bad_Part_S} shows the performance of the algorithm for this scenario. For a $\sim 50 \%$ smearing on the primary event position the fraction of mis-matches are found to be at the level of $10\%$, however, this represents a very conservative scenario for a realistic detector.

\section{\label{sec:PerformanceMultiS2}The delayed secondary signal matching problem with multiple scatter events}
A more important use case for application of matching may be when there are multiple secondary S2 signals as a result of more than one scattering of the single projectile particle in the detector. Complications may arise since there could be multiple S2 candidates corresponding to a single detected S1 signal, while the ranking table in the Stable Marriage problem only allows the first one to be matched to it. Therefore, the solution to the stable marriage problem fails on these type of events by construction. We carry on and study the College Admission algorithm which allows to match multiple S2 events to a single S1 event. For this case the S2 events can be considered as the students, the S1 events are the colleges and the scattering multiplicities associated with an S1 event are the colleges' quotas. 

We generate again toy Monte Carlo simulations but extend the event generation to allow for multiple scattering of S1 events and subsequent emission of multiple S2 events. The generated mean S1 event position is taken as the average of all the generated, random S1 positions from multiple scatterings, reflecting the fact that in LAr TPC detectors the primary scintillation events cannot be resolved for multiple scattering. The corresponding S2 events are generated on the hypothetical liquid scintillator top surface as before, assuming a fixed electron drift velocity. The scattering multiplicity of the generated S1 events is randomly sampled from a truncated Poisson distribution, where the zero event multiplicity is excluded, and the maximum allowed multiplicity considered was five scatterings. The algorithm is applied similarly to the Stable Marriage Problem example. The fundamental difference is the quota, which determines the maximum amount of S2 events allowed to be matched to a S1 event. However, for any event a priori the quota is unknown, since the scattering events occur randomly, and various quota values may be needed due to various processes with different multiplicities. We assume however that from the knowledge on the nature of the detector and the environment it is possible put an upper limit on the maximum number scatters that can occur in the detector volume. In the current work we put an arbitrarily large value of 10 in the algorithm as a quota, and let the algorithm match the S1 and S2 events following the Likelihood ordering rule, but filling up the quota by itself up to where the ranking and time constraints allow for. Therefore there is virtually no bias to any limit on the amount of matching that can occur to a S1 event. After the algorithm is finished on the reconstruction of generated events the multiplicity is determined by counting how many S2 events were associated with the each of the S1 events.

\begin{figure}[htb]
\centering
\includegraphics[height=0.65\textwidth]{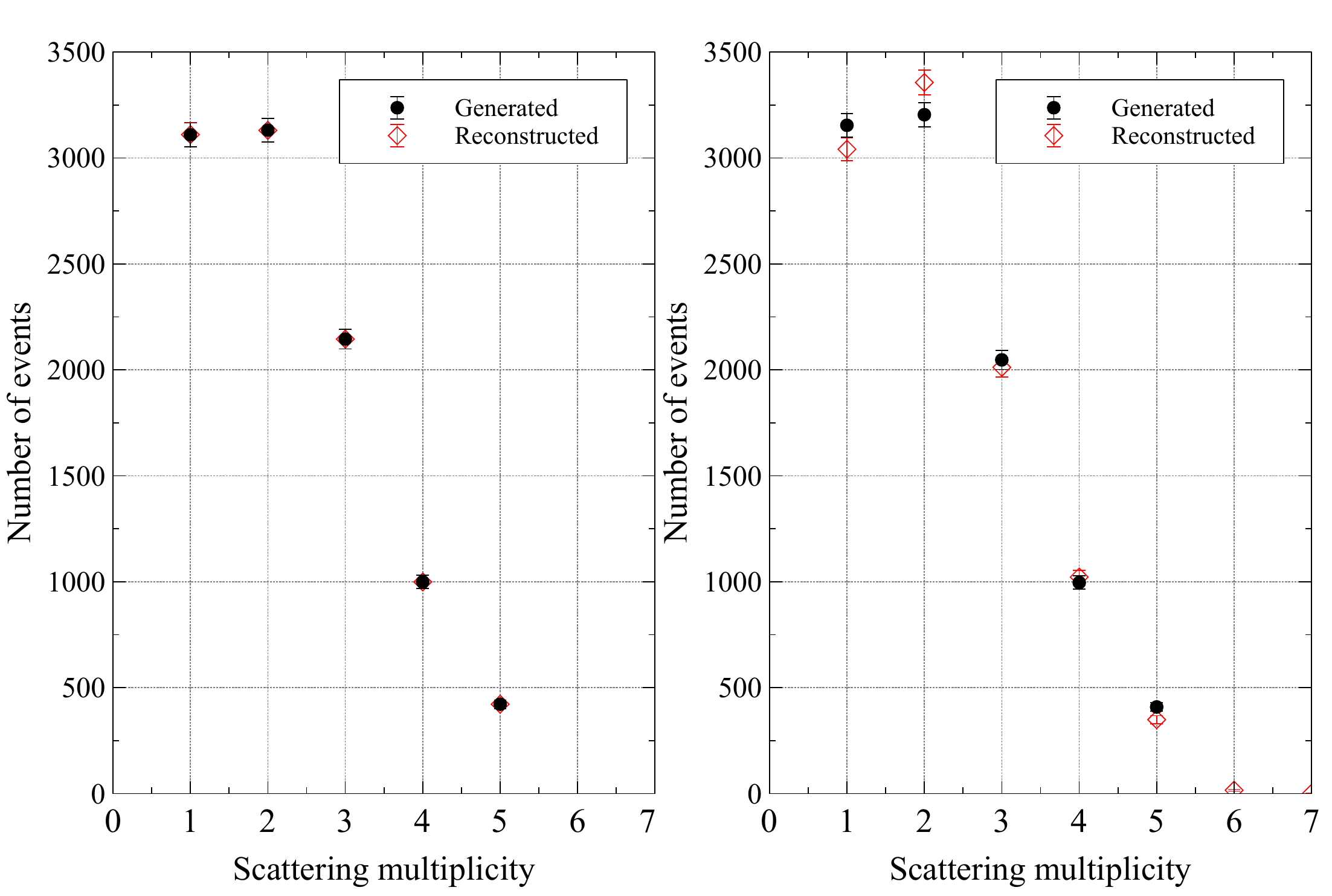}
\caption{Generated and reconstructed scattering multiplicity distribution for various values of the dimensionless variable, $\mathcal{M} = 0.57$ (left), $\mathcal{M} = 1.03$ (right), using the College Admission algorithm. }
\label{fig:CA_Mult_1}
\end{figure}

\begin{figure}[htb]
\centering
\includegraphics[height=0.65\textwidth]{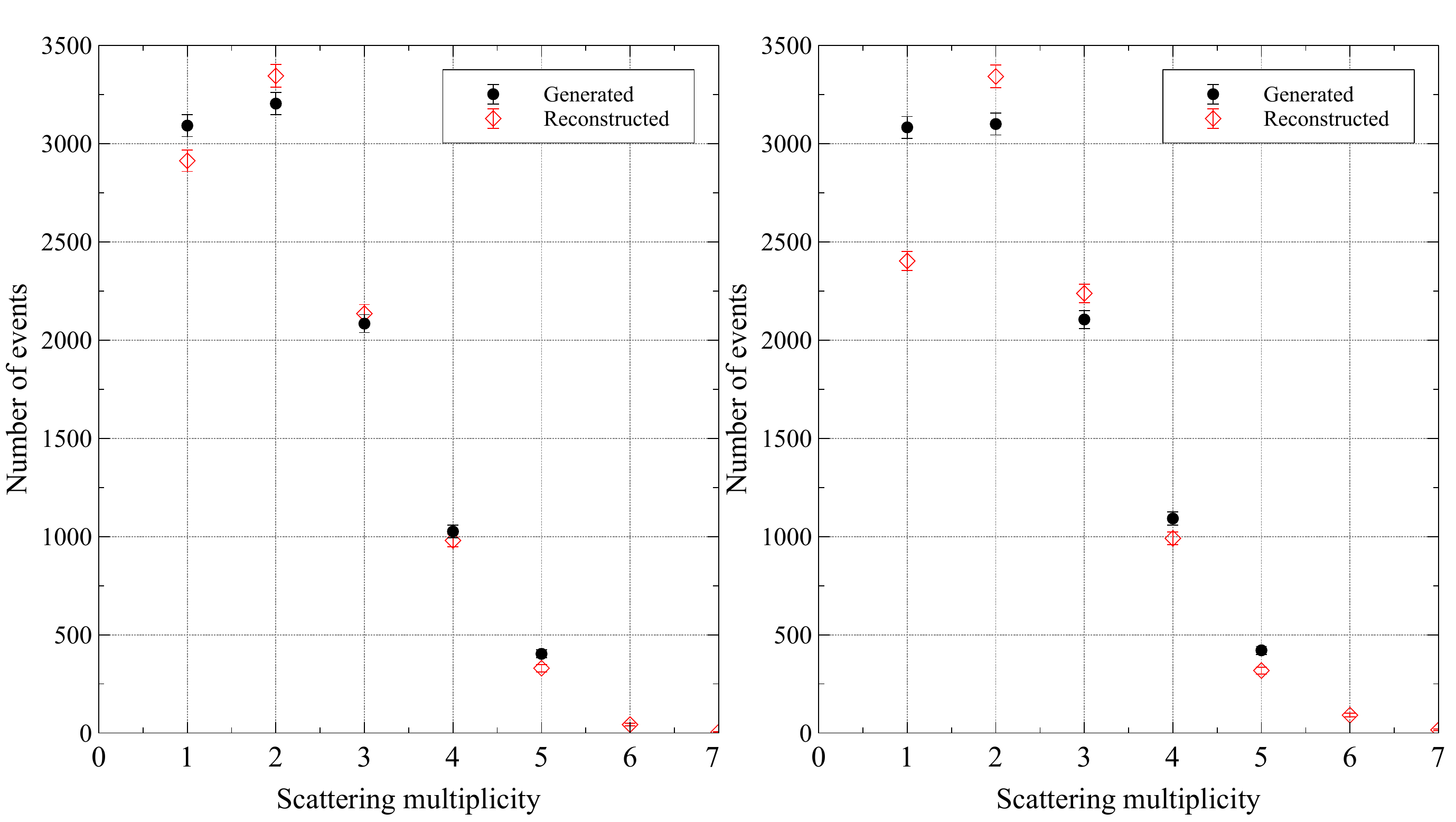}
\caption{Generated and reconstructed scattering multiplicity distribution for various values of the dimensionless variable,  $\mathcal{M} = 1.15$ (left),  $\mathcal{M} = 2.31$ (right), using the College Admission algorithm. }
\label{fig:CA_Mult_2}
\end{figure}

The reconstructed and true scattering multiplicity distributions are shown in Figure~\ref{fig:CA_Mult_1} and Figure~\ref{fig:CA_Mult_2} for various values of the dimensionless variable, $\mathcal{M}$.  The College Admission algorithm can reconstruct the multiplicity distribution remarkably well for $\mathcal{M} < 1$. Above this value the mis-matching rate increases, but depending the multiplicity of the events the amount of mis-matching varies. The events mostly affected are low scattering multiplicity events. For example for $\mathcal{M} = 2.3$ the mis-matching fraction for single scatter events is at the level of $\sim 22 \%$. With increasing scattering multiplicity the fraction of mis-matched events decreases, for scattering multiplicity of four the mis-match fraction is already at $\sim 2\%$. We have found that for the mis-matched events in case of $\mathcal{M} = 2.3$ the ranking algorithm systematically misses the true matching mostly by one rank in the Likelihood ranking table. This is clearly visible in Figure~\ref{fig:CA_Mult_2} as a systematic underestimation of the amount single scatter events and over estimation of the amount of double scatter events. This can be interpreted in a way that with additional experimental constraints the performance of the algorithm may be improved. Another result from the mis-matching is producing a small amount of events with reconstructed multiplicity larger than the maximal generated one. However, the fraction of these events is found to be negligible, less than $0.1\%$.
 
\begin{figure}[htb]
\centering
\includegraphics[height=0.65\textwidth]{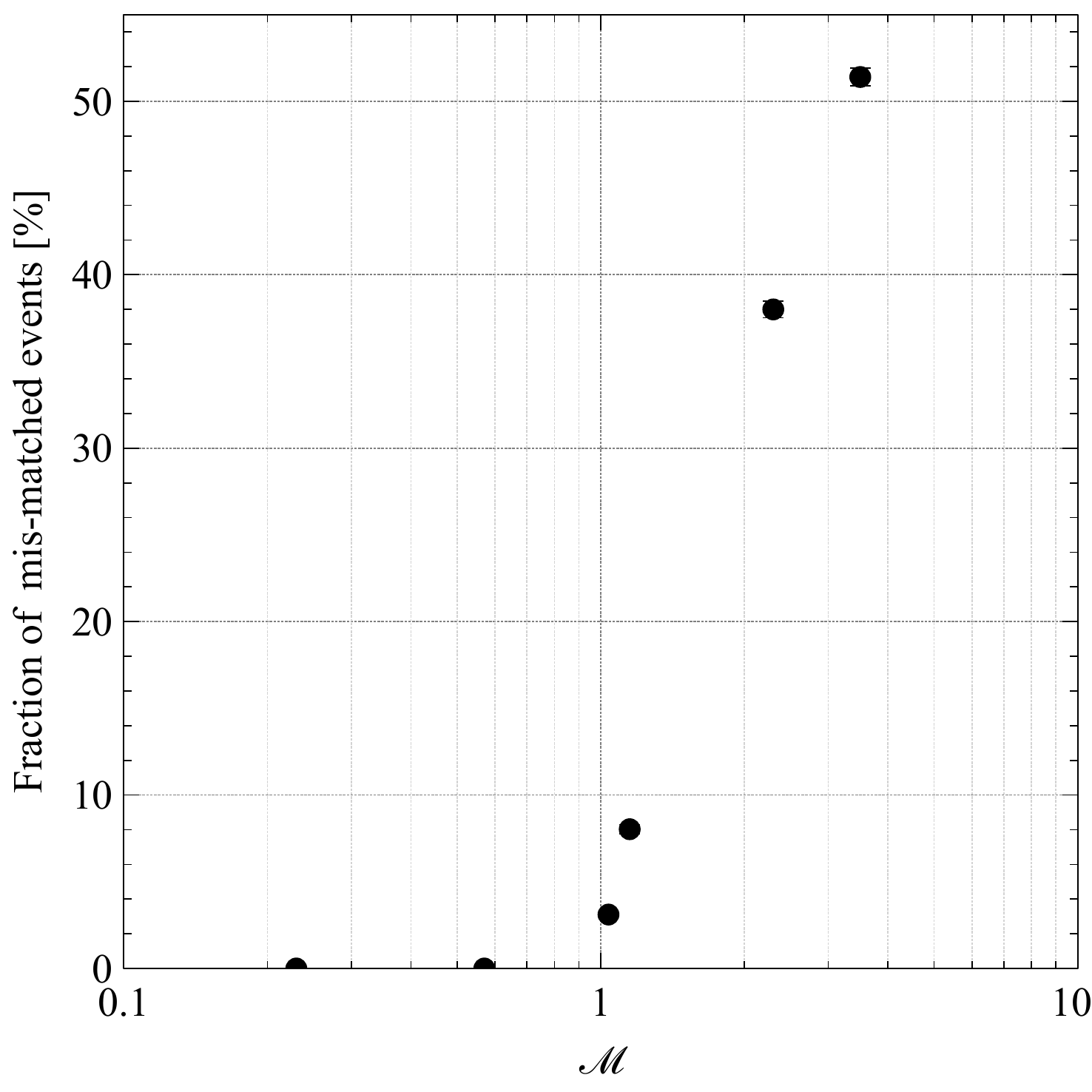}
\caption{Fraction of mis-matched events in the College Admission algorithm as a function of the dimensionless quantity $\cal{M}$. For each point $10k$ events were generated.}
\label{fig:CA_Bad}
\end{figure}

\begin{figure}[htb]
\centering
\includegraphics[height=0.7\textwidth]{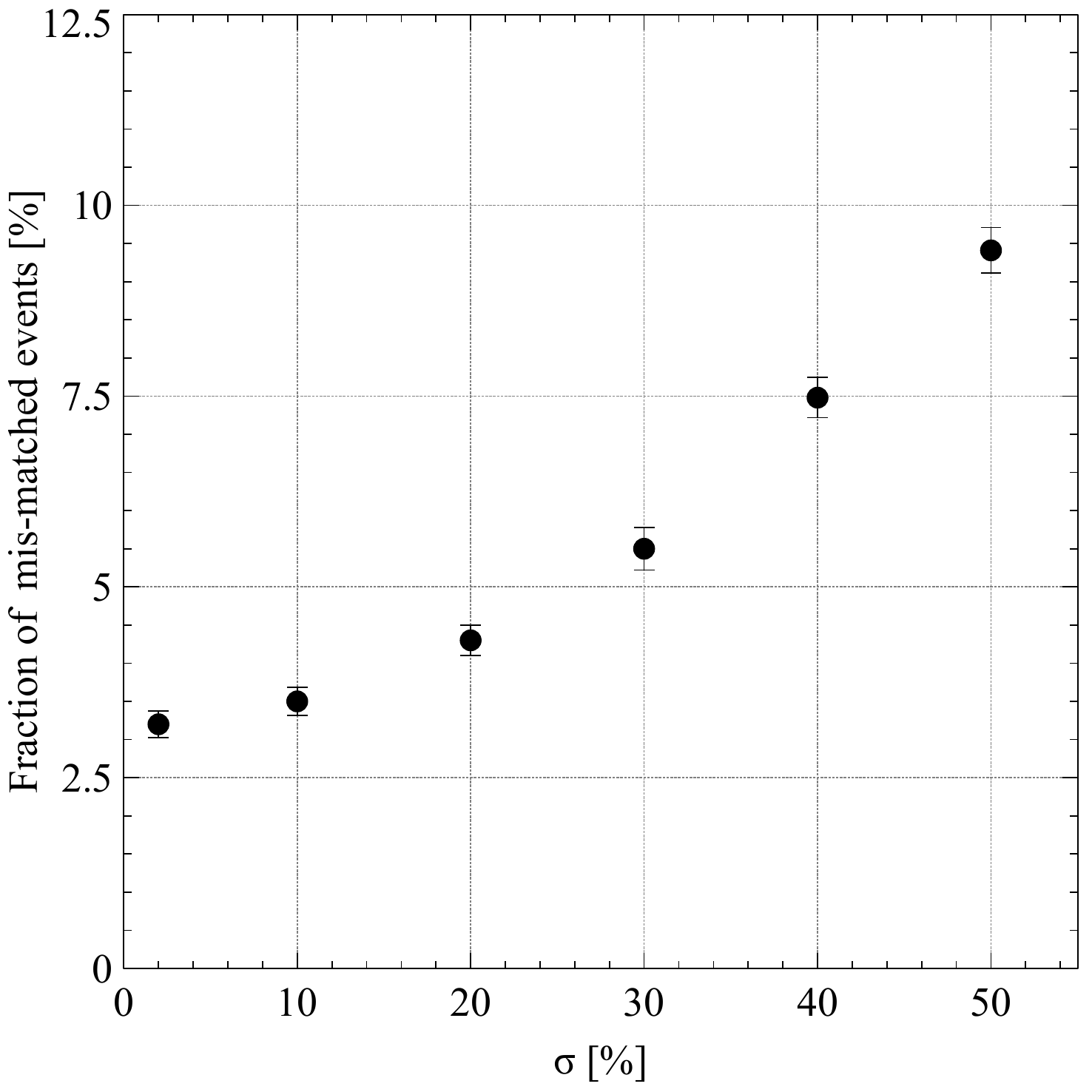}
\caption{Fraction of mis-matched events in the College Admission algorithm as a function of various relative amounts of Gaussian smearing, $\sigma$, on the average primary signal emission position, in units of percent, for events of class $\mathcal{M} = 1$. For each point $10k$ events were generated.}
\label{fig:CA_Smear}
\end{figure}

The average mis-matching fraction for the generated multiple scattering events is shown in Figure~\ref{fig:CA_Bad}. Similarly to the use case of the Stable Marriage algorithm, the mis-match fraction stays at below $\sim 1\%$ for $\mathcal{M} < 1$ , and increasing already to around $\sim 50\%$ at $\mathcal{M} \sim 3.5$. 
The effect of Gaussian smearing of the primary event position on the fraction of mis-matched events is slightly different, however, than for the Stable Marriage problem, it is shown in Figure~\ref{fig:CA_Smear}. For events of class $\mathcal{M} \simeq 1$ the mis-match fraction stays below $\sim 10 \%$. Considering the fact that the primary S1 events are already taken with the average of potentially multiple scattering positions, surprisingly, an additional smearing do not significantly change their ordering in the Likelihood ranking table. 

\begin{figure}[htb]
\centering
\includegraphics[height=0.6\textwidth]{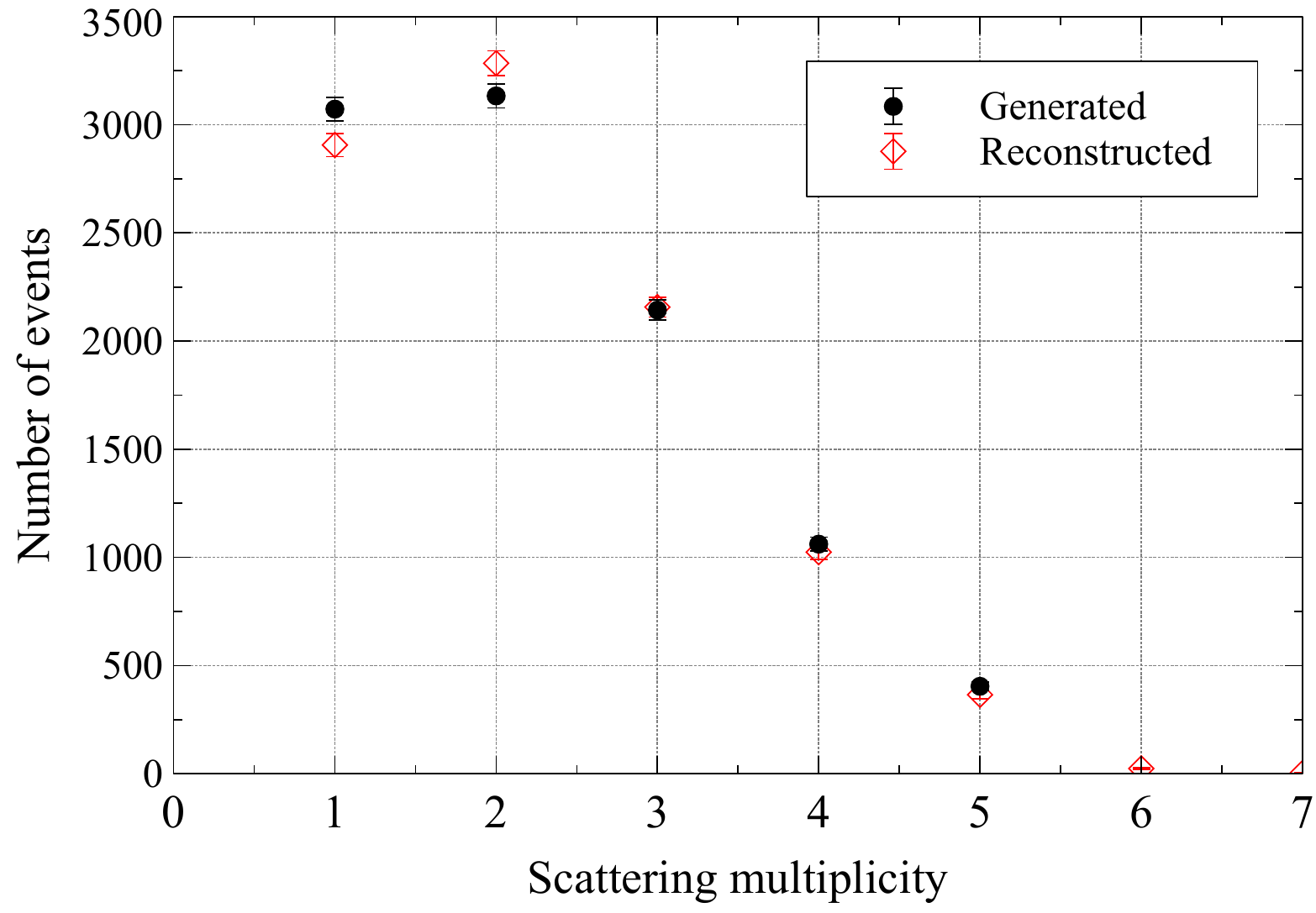}
\caption{Generated and reconstructed scattering multiplicity distribution for the case of $20\%$ smearing on the S1 position, and for class of $\mathcal{M} = 1$ events, using the College Admission algorithm.}
\label{fig:CA_multi_Smear}
\end{figure}

As an example the generated and reconstructed multiplicity distribution is shown for the case of $20\%$ Gaussian smearing of the average S1 position for $\mathcal{M} = 1$ events in Figure~\ref{fig:CA_multi_Smear}. The multiplicity distribution has been recovered well from the data by the algorithm.

It is noted that in both algorithms discussed, the various stages of the event processing (evaluation of the Likelihood function, filling up the ranking tables, and performing the iterative matching algorithms) do not require significant processing time. In a few iterations the matching algorithm finishes, and each iteration only requires comparing and sorting of a few scalar numbers (depending on the magnitude of the $\mathcal{M}$ parameter) and storage of pairs of event IDs. Therefore in general both Gale-Shapely algorithms are computationally inexpensive.

\section{\label{sec:Conclusion}Conclusions}
A framework is presented for the problem of matching primary and delayed secondary signals, naturally occurring in DP LAr TPC detectors used for single or multiple scatter events. In the context of the Gale-Shapley solution to the stable marriage problem a Likelihood-based ordering rule is proposed for ranking the probability of agreement between various primary and secondary signal candidate pairs. The parametrisation of the Likelihood function depends on the subdetector-level information available. With a perfectly understood detector such tuning of the parameters of the ordering Likelihood function allows to optimize the matching procedure. For single scatter events in the simplified toy Monte Carlo example, the results suggest reasonably good performance. The algorithm was capable to perform correct match making with a fraction of mis-matches at $1 \%$ level, even at extreme values of event rates. Introducing various levels of position smearing the mis-matched fraction may reach the level of $\sim 10 \%$. For multiple scatter events the College Admission algorithm is applied, allowing multiple secondary events to be matched to a single primary event. The matching algorithm can successfully reconstruct the scattering multiplicity in randomly generated multiplicity values for $\mathcal{M} \leq 1$ class of events, and the mis-matching fraction stays below a few percent. In particular, for large scale, low background liquid scintillator Dark Matter search experiments the rate of primary events and delay time of secondary events fall into a range of class $\mathcal{M} < 1$, therefore the Gale-Shapley algorithms are very prospective candidates for event reconstruction. The performance of the algorithms on data will be presented in a separate publication.

\section{Acknowledgments}

This work was supported by the Swiss National Science Foundation (SNF) Grant $200020$
\_$162794$.\\

\section*{References}
\bibliography{Bibliography.bib}

\end{document}